\def\gore{M.\,TOMCZAK, E.\,CHMIELEWSKA: XPEs and other solar-activity phenomena}

\documentclass{ceab}   

\usepackage{epsfig}     
\usepackage{graphicx}   

\usepackage{ceabbib}     
\usepackage[T1]{fontenc}

\begin{document}

\title{X-ray Plasma Ejections and their association with other solar-activity phenomena}

\author{\normalsize M.\,TOMCZAK, E.\,CHMIELEWSKA \vspace{2mm} \\
        \it Astronomical Institute, University of Wroc{\l }aw, \\
        \it  ul.\,Kopernika 11, PL-51-622 Wroc{\l }aw, Poland}

\maketitle

\begin{abstract}
Recently developed in Wroc{\l }aw, a new catalogue of X-ray Plasma
Ejections (XPEs) observed by the Soft X-ray Telescope onboard {\sl
Yohkoh}, is a very useful tool for some statistical studies. We
investigated the association of XPEs with solar flares and coronal
mass ejections (CMEs). We found that particular subclasses of XPEs
show different levels of association. Moreover, flares and CMEs
associated with different subclasses of XPEs show distinctly
different characteristics. We conclude that the event that we call
'X-ray Plasma Ejection' can be a manifestation of different physical
processes.
\end{abstract}


\section{Introduction}

X-ray plasma ejections (XPEs) are sudden expulsions of hot
magnetized plasma in the solar corona seen in X-rays. XPEs display a
wide range of macroscopic motions showing different morphological,
kinematic, and physical conditions. They occur usually during the
impulsive phase of flares. Sometimes a given flare produces even
more ejections at later times and different locations. XPEs also
show a close connection with other solar-activity phenomena: coronal
mass ejections (CMEs), prominences, radio bursts, coronal dimmings,
and global waves.

XPEs have been systematically observed since 1991 when the {\sl
Yohkoh} satellite began operations (Klimchuk {\it et al.}, 1994,
Shibata {\it et al.}, 1995), nevertheless some earlier observations
are known (e.g., Harrison {\it et al.}, 1985). A strong
inhomogeneity of XPEs as a group suggests different physical
mechanisms (magnetic reconnection, magnetic loss-of-equilibrium,
others?) responsible for their occurrence.

Although other recent imaging instruments: the Soft X-ray Imager
onboard {\sl Geostationary Operational Enviromental Satellites (GOES)}, the
{\sl Reu-ven Ramaty High-Energy Solar Spectroscopic Imager (RHESSI)},
the X-Ray Telescope onboard {\sl Hinode}, have been providing
observations of XPEs, the Soft X--ray Telescope, SXT, (Tsuneta {\it
et al.}, 1991) onboard {\sl Yohkoh} resulted in the largest database
of XPEs until now.

\section{The YOHKOH/SXT XPE Catalogue}

We recently developed a new catalogue of XPs at the Astronomical
Institute of the University of Wroc{\l }aw. The catalogue contains
records of 368 XPEs observed by the {\sl Yohkoh}/SXT during the full
satellite operation (1991-2001). 163 events out of the 368 were not
reported before. The catalogue resides online at
http://www.astro.uni.wroc.pl/XPE/catalogue.html.

Each record in the catalogue contains MPEG movies illustrating
evolution of a XPE, a short qualitative description, results of
quantitative analysis (if available), references, and entries to the
associated flare and CME. In the case of flares and CMEs we used
data from the {\it Yohkoh} Flare Catalogue (HXT/SXT/SXS/HXS) -- Sato
{\it et al.} (2006) and the {\it SOHO} LASCO CME Catalog (Gopalswamy
{\it et al.}, 2009).

In our catalogue we developed a new classification scheme of XPEs
based on three criteria concerning: (a) the morphology of the XPE,
(b) its kinematics, and (c) multiplicity of the occurrence.
Examining each criterion we distinguish two subclasses of events
only: (a) 1 -- collimated, 2 -- loop-like; (b) 1 -- confined, 2 --
eruptive; (c) 1 -- single, 2 -- recurrent. In each criterion the
subclass 1 refers less energetic events and the subclass 2 refers
more energetic events.

The morphological criterion resolves the direction of the moving
soft X-ray plasma in comparison with the local magnetic field. In
the case of the subclass 1, the direction is parallel, i.e. along
the already existing field lines; in the case of the subclass 2 ---
perpendicular, i.e. across the already existing field lines (or
strictly speaking -- together with them).

For the assignment into one of the kinematical subclasses we have
chosen a rate of the height increment above the chromosphere,
$\dot{h}$. A negative value, $\dot{h} < 0$, means the subclass 1,
the opposite case, $\dot{h} \ge 0$ means the subclass 2. XPEs from
the kinematical subclass 1 suggest the presence of magnetic or
gravitational confinement. For XPEs from the kinematical subclass 2,
an increasing velocity in the radial direction in the field of view
of the SXT allows us to anticipate a further expansion leading to
irreversible changes (eruption) of the local magnetic field. In
consequence, at least a part of the plasma escapes from the Sun.

According to our third criterion we separate unique XPEs that
occurred once in time (subclass 1) from recurrent events for which
following expanding structures can be seen with time (subclass 2).

By using the classification we were able to separate several
subclasses of XPEs that look more homogeneous than the full
population. Unfortunately, we are not sure if the particular
subclasses of XPEs refer events that are physically different. A
closer confirmation would give a quantitative analysis of the soft
X-ray images, however for the majority of XPEs in the catalogue this
kind of analysis is practically unreliable, due to minor signal and
other observational limits. Therefore, the main motivation of our
paper is to justify the presence of physically different subclasses
of XPEs based on a comparison of the properties of other associated
solar-activity phenomena. For events like flares or CMEs their basic
characteristics are known well. In Section 3 we present the
association of XPEs with flares, in Section 4 we present the
association of XPEs with CMEs.

\section{Association of XPEs with flares}

Recognition of an associated flare for the majority of XPEs in the
catalogue is very easy. The flare is seen usually in movies
illustrating evolution of the XPE as a strong saturation due to much
stronger soft X-ray radiation. In several cases a flare occurred
simultaneously with the XPE but in an other active region. We
considered this flare as the associated event only in case when some
distinct magnetic loops connecting both active regions were seen.
Finally, there are several XPEs in the catalogue that occurred when
no flare was observed on the Sun.

For the associated flares we determined the X-ray class and the
total duration basing on light curves recorded by {\sl GOES}, in the
wavelength interval 1-8 \AA\ . We defined the total duration as the
interval between a constant level of the solar flux before and after
the flare, therefore our values of this parameter are larger than
intervals between a start time and end time that are routinely
reported in {\sl Solar-Geophysical Data}.

\begin{figure}[t]
\epsfig{file=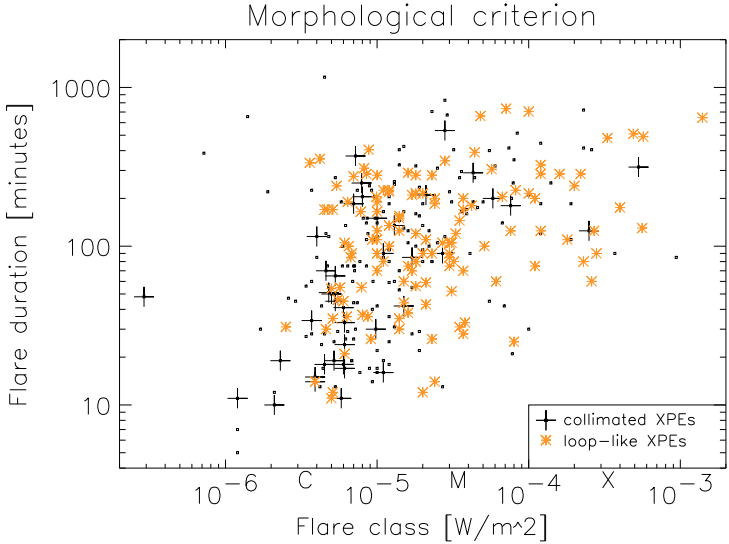,width=10.5cm} \caption{Scatterplot of
the flare X-ray class versus the flare total duration. Flares
associated with the subclasses of collimated and loop-like XPEs are
marked with crosses and stars, respectively. Small boxes refer to flares
associated with XPEs observed with a poor quality or during other
problematic circumstances.}
\end{figure}

In Fig.\,1 we present a scatterplot of the X-ray class versus the
total duration for flares associated with the morphological
subclasses of XPEs, i.e., collimated and loop-like XPEs marked with
boxes and stars, respectively. The aggregate number of events is far
lower than the total number of events in the catalogue, because we
considered only well-observed XPEs and flares that are non-occulted
by the solar disk. Both groups of flares are mixed in the plot,
however some shift toward higher X-ray class and longer duration is
seen for flares associated with loop-like XPEs.

This shift is confirmed by medians calculated separately for both
groups of flares. As it is seen in Table 1, the median X-ray class
of flares associated with loop-like XPEs is three times more than
that for collimated XPEs. Similarly, the median flare duration of
loop-like XPEs is two times greater than that of the collimated
XPEs. Higher X-ray class and longer duration mean a more-energetic
flare. Thus we can conclude that more-energetic, loop-like XPEs are
associated with more-energetic flares, on average, and the
less-energetic, collimated XPEs rather prefer less-energetic flares.

\begin{table}[t]
\caption{Properties of flares associated with particular subclasses
of XPEs}
\begin{center}
\begin{tabular}{llccc}
\hline
 Classification & XPE & Number & Flare & Flare \\
 criterion & subclass & of events & class & duration \\
 & & & (median) & (median) \\
\hline \hline I (morphological): & & & & \\
 & 1 (collimated) & 40 & C6.1 & 51 min. \\
 & 2 (loop-like) & 126 & M1.8 & 110 min. \\
 & & & {\bf x3.0} & {\bf x2.2} \\
 II (kinematical): & & & & \\
 & 1 (confined) & 49 & C6.1 & 45 min. \\
 & 2 (eruptive) & 94 & M2.3 & 120 min. \\
 & & & {\bf x3.8} & {\bf x2.7} \\
 III (recurrence): & & & & \\
 & 1 (single) & 53 & M1.4 & 75 min. \\
 & 2 (recurrent) & 65 & M2.1 & 155 min. \\
 & & & {\bf x1.5} & {\bf x2.1} \\
 & & & & \\
 & 1+1+1 & 13 & C5.2 & 42 min. \\
 & 2+2+2 & 41 & M3.1 & 155 min. \\
 & & & {\bf x6.0} & {\bf x3.7} \\
 \hline
\end{tabular}
\end{center}
\end{table}

We investigated also the characteristics of flares associated with
subclasses of XPEs defined by our kinematical and recurrence
criteria. Medians calculated separately for flares associated with
particular subclasses of XPEs, presented in Table 1, show a similar
tendency, namely that the association between XPEs and flares is
determined by amount of energy that were released in these events.
The difference between values of median is higher for flares
associated with subclasses of XPEs defined by the kinematical
criterion (a factor 3.8 and 2.7 for X-ray class and duration,
respectively) than for flares associated with subclasses defined by
the recurrence criterion (a factor 1.5 and 2.1 for X-ray class and
duration, respectively).

We can expect that a difference between the characteristics that
describe associated flares should be even higher for two subclasses
of XPEs that we define by combining our three criteria
simultaneously. Indeed, medians for flares associated with the
subclass 1+1+1 (collimated, confined, single XPEs) and subclass
2+2+2 (loop-like,eruptive, recurrent XPEs) show extreme differences
(a factor 6.0 and 3.7 for X-ray class and duration, respectively).
As it is seen in Fig.\,2, the flares associated with the subclass
1+1+1 and 2+2+2 of XPEs are almost separated.

\begin{figure}[t]
\epsfig{file=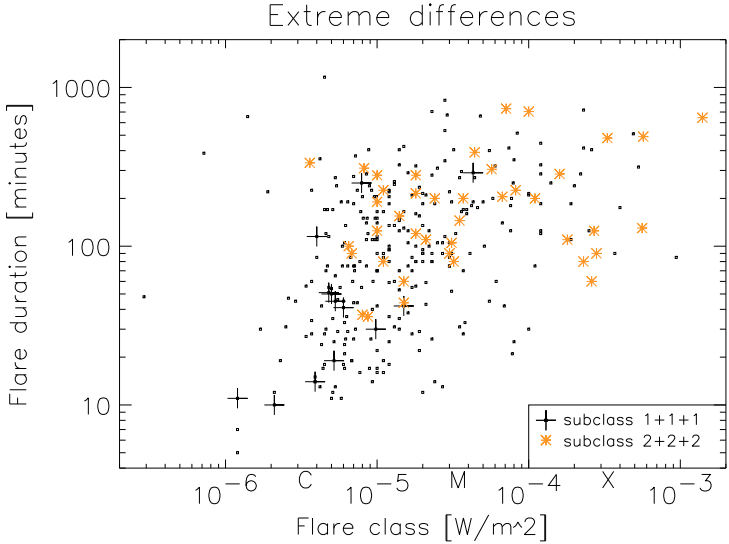,width=10.5cm} \caption{Scatterplot of
the flare X-ray class versus the flare total duration. Flares
associated with the subclasses of collimated, confined, single
(1+1+1) and loop-like, eruptive, recurrent (2+2+2) XPEs are marked
with crosses and stars, respectively. Small boxes refer to flares associated
with other XPEs.}
\end{figure}

\section{Association of XPEs with CMEs}

\begin{figure}[t]
\epsfig{file=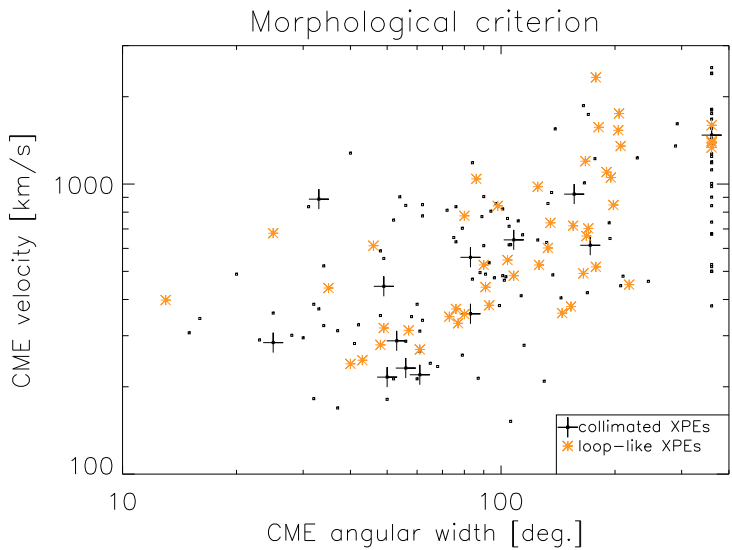,width=10.5cm} \caption{Scatterplot of the CME
angular width versus the CME linear velocity. CMEs associated with the
subclasses of collimated and loop-like XPEs are marked with crosses
and stars, respectively. Values are taken from the SOHO LASCO CME
catalog (Gopalswamy {\it et al.}, 2009). Small boxes refer to CMEs associated
with XPEs observed with a poor quality or during other problematic
circumstances.}
\end{figure}

For associating the XPEs from our list with CMEs we used the {\sl SOHO}
LASCO CME Catalog (Gopalswamy {\it et al.}, 2009). Only 296 XPEs
occurred when the LASCO coronagraphs were making observations. We
found that 180 XPEs (60.8\%) were associated with CMEs. This is
slightly lower than the 69\% (95 from 137 events) obtained by Kim {\it
et al.} (2005). We consider a XPE-CME pair as physically connected
if the XPE occurred within the position angles defined by the CME
angular width increased by 10$^{\circ}$ from both sides. Moreover,
time of XPE occurrence had to fall within 3-hours-interval centered
around the extrapolated time of CME start at $h=1 R_S$. For the
extrapolation we used the time of first appearance in the LASCO/C2 field
of view and the linear velocity taken from the CME catalog.

In Fig.\,3 we present scatterplot of angular width versus linear
velocity for CMEs associated with morphological subclasses of XPEs,
i.e., collimated and loop-like XPEs marked with boxes and stars,
respectively. The aggregate number of events is far lower than the
total number of events in the catalogue, because we considered only
well-observed XPEs that occurred close to the solar limb ($|\lambda|
> 60^{\circ}$). Both groups of CMEs are mixed in the plot, however
some shift toward wider and faster events is seen for CMEs
associated with loop-like XPEs.

\begin{table}[t]
\caption{Properties of CMEs associated with particular subclasses of
XPEs}
\begin{center}
\begin{tabular}{llccc}
\hline
  & & & CME & CME \\
 Classification & XPE & Number & angular & velocity \\
 criterion & subclass & of events & width & (median) \\
 & & & (median) & [km/s] \\
\hline \hline I (morphological): & & & & \\
 & 1 (collimated) & 13/31 (42\%) & 61$^{\circ}$ & 444 \\
 & 2 (loop-like) & 47/71 (66\%) & 126$^{\circ}$ & 602 \\
 & & & {\bf x2.1} & {\bf x1.4} \\
 II (kinematical): & & & & \\
 & 1 (confined) & 16/41 (39\%) & 83$^{\circ}$ & 526 \\
 & 2 (eruptive) & 37/48 (77\%) & 126$^{\circ}$ & 642 \\
 & & & {\bf x1.5} & {\bf x1.2} \\
 III (recurrence): & & & & \\
 & 1 (single) & 17/31 (55\%) & 104$^{\circ}$ & 518 \\
 & 2 (recurrent) & 24/33 (73\%) & 126$^{\circ}$ & 602 \\
 & & & {\bf x1.2} & {\bf x1.2} \\
 & & & & \\
 & 1+1+1 & 7/14 (50\%) & 61$^{\circ}$ & 357 \\
 & 2+2+2 & 20/23 (87\%) & 126$^{\circ}$ & 613 \\
 & & & {\bf x2.1} & {\bf x1.7} \\
 \hline
\end{tabular}
\end{center}
\end{table}

This shift is confirmed by medians calculated separately for both
groups of CMEs. As it is seen in Table 2, the median for CMEs
associated with loop-like XPEs is 2.1 times and 1.4 times greater
than the median for CMEs associated with collimated XPEs for CME
angular width and CME velocity, respectively. Higher angular width
and velocity mean a more-energetic CME, thus we can conclude that
more-energetic, loop-like XPEs are associated with more-energetic
CMEs, on average, and the less-energetic, collimated XPEs rather prefer
the less-energetic CMEs. Moreover, the loop-like XPEs show better correlation
with CMEs than the collimated ones: 66\% and 42\%, respectively.

We also investigated characteristics of CMEs associated with
subclasses of XPEs defined by our kinematical and recurrence
criteria. Medians calculated separately for CMEs associated with
particular subclasses of XPEs, presented in Table 2, show a similar
tendency, namely that the association between XPEs and CMEs is
determined by the amount of energy released in these events.
The difference between the median values is slightly higher for CMEs
associated with subclasses of XPEs defined by the kinematical
criterion (a factor 1.5 and 1.2 for angular width and velocity,
respectively) than for CMEs associated with subclasses defined by
the recurrence criterion (a factor 1.2 for both angular width and
velocity). More-energetic XPEs better correlate with CMEs than
less-energetic subclasses: 77\% and 39\%, respectively, for
kinematical criterion, and 73\% and 55\%, respectively, for
recurrence criterion.

As in case of flares, medians for CMEs associated with the subclass
1+1+1 (collimated, confined, single XPEs) and subclass 2+2+2
(loop-like, eruptive, recurrent XPEs) show extreme differences (a
factor 2.1 and 1.7 for angular width and velocity, respectively).
The subclass 2+2+2 shows the strongest correlation with CMEs (87\%),
whereas the subclass 1+1+1 -- only 50\%.

\section{Discussion and Conclusions}

Our investigation shows that the characteristics of flares and CMEs
associated with particular subclasses of XPEs are different. We
found that the scale of differences is higher for flares than for
CMEs. We also found that the morphological and kinematical criteria proposed in our classification scheme of XPEs separate better the associated events
than the recurrence criterion (compare the boldfaced lines in Tables 1 and 2).

The results strongly suggest
that the total amount of energy, converted from the magnetic field in
the active region during its magnetic reconfiguration, determines the
characteristics of events like: flares, CMEs, XPEs, which are common
consequences of this reconfiguration. Thus, more-energetic XPEs are
associated with more-energetic flares and CMEs and less-energetic ones
-- seem to occur commonly. This statistically averaged picture has exceptions in the partitioning of the magnetic energy.

We have shown that the subclasses of XPEs separated on the basis of
our simple observational criteria have different levels of
correlation with other solar-activity phenomena. The difference is
also seen if we consider the basic parameters describing these flares
and CMEs. However, the association of XPEs with different flares or
CMEs does not mean a specific eruption mechanism as long as these
flares or CMEs do not represent physically different groups. The everlasting discussion concerning the reality of existence of two different groups of flares (compact vs. arcade, e.g., Pallavicini {\it et al.}, 1977) or CMEs (accelerated vs. constant, e.g., Andrews \& Howard, 2001) is crucial in this context.

A careful inspection of many movies in the XPE catalogue suggests
different solutions. For example, the collimated XPEs seems to be
connected more directly with the reconnection process (reconnection
outflow, chromospheric evaporation), whereas the loop-like XPEs are
connected rather with a loss-of-equilibrium of magnetic structures.
In some examples occurs also a leakage of plasma due to the plasma-$\beta$
parameter approaching unity.

For a more precise separation of physically different subclasses of
XPEs a quantitative analysis of the XPE observations is necessary (see
Tomczak \& Ronowicz, 2007).

\section*{Acknowledgements}
The {\sl Yohkoh} satellite is a project of the Institute of Space
and Astronautical Science of Japan. The CME catalog used in this
work is generated and maintained at the CDAW Data Center by NASA and
The Catholic University of America in cooperation with the Naval
Research Laboratory. We acknowledge useful comments and plenty of corrections made by the Referee, Dr. N. Gopalswamy. This work was supported by a Polish Ministry of
Science and High Education grant No. N\,N203\,1937\,33.


\section*{References}
\begin{itemize}
\small
\itemsep -2pt
\itemindent -20pt
\item[] Andrews, M.\,D., Howard, R.\,D.: 2001, {\it Space Sci. Rev.}, {\bf 95}, 147.
\item[] Gopalswamy, N., Yashiro, S., Micha{\l }ek, G., {\it et al.}: 2009,
{\it Earth Moon Planet}, {\bf 104}, 295.
\item[] Harrison,\,R.\,A., Waggett,\,P.\,W., Bentley,\,R.\,D., {\it et al.}:
1985, {\it Solar\,Phys.}, {\bf 97}, 387.
\item[] Kim,\,Y.-H., Moon,\,Y.-J., Cho,\,K.-S., {\it et al.}: 2005,
{\it Astrophys.\,J.}, {\bf 622}, 1240.
\item[] Klimchuk,\,J.\,A., Acton,\,L.\,W., Harvey,\,K.\,L., {\it et
al.}: 1994, {\it in} Y.\,Uchida, T.\,Watana-be, K.\,Shibata,
H.\,S.\,Hudson (eds.) {\it X--ray solar physics from {\sl Yohkoh}},
Universal Academy Press, 181.
\item[] Pallavicini, R., Serio, S., Vaiana, G.\,S.: 1977, {\it Astrophys.\,J.}, {\bf 216}, 108.
\item[] Sato, J., Matsumoto, Y., Yoshimura, K., {\it et al.}: 2006, {\it Solar Phys.},
{\bf 236}, 351.
\item[] Shibata,\,K., Masuda,\,S., Shimojo,\,M., {\it et al.}: 1995,
{\it Astrophys.\,J.\,Lett.}, {\bf 451}, L83.
\item[] Tomczak, M., Ronowicz, P.: 2007, {\it Cent. Eur. Astrophys.
Bull.}, {\bf 31}, 115.
\item[] Tsuneta,\,S., Acton,\,L., Bruner,\,M., {\it et al.}: 1991,
{\it Solar Phys.}, {\bf 136}, 37.
\end{itemize}

\bibliographystyle{ceab}
\bibliography{sample}

\end{document}